\documentclass[12pt,preprint]{aastex}

\begin{document}

\title{Influence of an Internal Magnetar on Supernova Remnant Expansion}

\author{M. P. Allen and J. E. Horvath}
\affil{Instituto de Astronomia, Geof\'\i sica e Ci\^encias Atmosf\'ericas,\\
Universidade de S\~ao Paulo \\ R. do Mat\~ao 1226 - Cidade Universit\'aria\\
(05508-900) S\~ao Paulo SP - Brazil}

\email{mpallen@astro.iag.usp.br}

\begin{abstract}
Most of the proposed associations between magnetars and supernova remnant
suffer from age problems. Usually, supernova remnants ages are 
determined from an approximation of the Sedov-Taylor phase relation between 
radius and age, for a fixed energy of the explosion $\sim 10^{51}$ 
erg. Those ages do not generally agree with the characteristic ages of the (proposed) associated 
magnetars. We show quantitatively that, by taking into account the energy injected on the supernova 
remnant by magnetar spin-down, a faster expansion results, improving matches between
characteristic ages and supernova remnants ages.
However, the magnetar velocities inferred from observations would inviabilize some associations.
Since characteristic ages may not be good age estimators, their influence on the
likelihood of the association may not be as important.

In this work we present simple numerical simulations of supernova remnants 
expansion with internal magnetars, and apply it to the observed objects. 
A short initial spin period, thought to be important for the very generation of
the magnetic field, is also relevant for the modified expansion of the remnant.
We next analyze all proposed associations case-by-case, addressing the likelyhood 
of each one, according to this perspective. We consider a larger explosion
energy and reasses the characteristic age issue, and conclude that
about 50\% of the associations can be true ones, provided SGRs and AXPs are magnetars.

\end{abstract}


\section{Introduction}

Soft Gamma Repeaters (SGRs) and Anomalous X-ray Pulsars (AXPs) are two classes of
objects being candidates to magnetars, neutron stars with magnetic fields above the
quantum critical threshold $B_c \, = \, 4.41 \times 10^{13}$ G. In the model proposed
by \citet{mag}, which we will follow through this work, dynamo action in a fast 
rotating proto-neutron star amplifies the ``seed'' magnetic field ($10^{11} -
10^{13}$ G) by one to three orders of magnitude. The simplest estimate of the magnetic field
intensity is derived from measurements of their rotational periods ($P$) and period
derivatives ($\dot P$), through the well known expression

\begin{equation}
B \sin \chi \, = \, {\left( {{3 c^3 I P \dot P} \over {8 \pi R_{psr}^6}}\right)}^{1/2} \; ,
\end{equation}

\noindent
where $B$ is the magnetic field intensity, $R_{psr}$ is the pulsar radius ($\sim
10^6$ cm), $\chi$ is the angle between the rotational and magnetic field axis, $I$ is the moment
of inertia ($\sim 10^{45}$ g cm$^2$), and $c$ is the speed of light. Their ages can also
be estimated through the characteristic age

\begin{equation}
\tau \, = \, {P \over {2 \dot P}} \; ,
\end{equation}

\noindent
but both calculations assume continuous spin-down driven by magnetic dipole
braking, an assumption which has been questioned by several authors
\citep{nn,venrel,PTH95,CHN00,kou99}, in different
contexts, regarding AXPs and SGRs. The derived figures are shown in Table 1. As
most of the objects are younger than 10$^4$ years, it seems reasonable to look
for the  Supernova Remnant (SNR) that was originated at the same supernova
explosion. In fact, almost all SGRs and AXPs were tentatively associated to
some SNR. Though, the likelihood of those associations is put in doubt because
the age estimated for those SNRs is generally larger than the characteristic
age of the associated magnetar.

This situation leads to the ``age'' and ``velocity'' problems of SGRs and AXPs. Both problems 
are related to the relatively low characteristic ages found for these 
objects. The SNRs usually associated to them seem to 
be systematically older, thus the age problem. Some SGRs and AXPs have been found to
lie beyond, at or near the border of their proposed SNRs, and, given their low ages, the implied velocities 
are much higher than those found for ordinary pulsars (see \citealp{sgraxp} for
a discussion about these objects). As seen on Tables 1 and 2, there is 
a large uncertainty in those estimates, mainly due to uncertainties in the
distances. This implies uncertainties in the radii on which age estimates are
based through models of the expansion.

Several authors have addressed magnetar-SNR associations. Some of them have
discarded either the magnetar hypothesis or the association, based on those
problems. This became known as the ``nature versus nurture'' debate, where one side
proposes that the characteristics of SGRs and AXPs are derived from their special nature
(see \citealt{axp}), while the other side finds more plausible that the 
environment is the actual responsible, nurturing the otherwise ordinary neutron stars 
with a fossil accretion disk \citep{nn,CHN00,PTH95}.

Associations between pulsars and SNRs are generally examined according to the following
criteria: 1) positional coincidence in sky; 2) distance
estimates; 3) age estimates; 4) evidences for interaction between the neutron star and
the SNR. Additional criteria may be derived from these, like the estimate of the
projected velocity on sky of the compact object. \citet{gae01} did a comprehensive
study of positional coincidences for magnetar-SNR associations. Distance can be
determined by neutral hydrogen column density measurements, evidence of association
with another object (e. g., HII region or molecular cloud) whose distance is already
known, or through surface brightness-diameter ($\Sigma -D$) relations. This last method 
is not considered reliable enough, because it ignores the effect of density and 
structure of the local interstellar medium, resulting in a large dispersion of values 
from the best fit, as can be seen in \citet{sigd}. Some SNRs on the proposed
associations show plerions, usually taken as positive signal of the presence of a
neutron star, but most do not present that feature. Finally, measurements of
HI column density are subject to systematic errors from models, and usually are not
better than 50\%, implying on similar uncertainties in distance determinations. Radio
pulsars can have their distance estimated by use of dispersion measure, however SGRs
and AXPs have not yet been detected in radio waves. There are reasons to believe they 
cannot emmit radio signals, either because they are magnetars, where the huge magnetic
field induces photon splitting rather than pair creation (\citealp{split}, but see
\citealt{Bgrande}), or in the case they are accretors, because accretion
quenches radio emission \citep{CHN00,PTH95}.

SNR age estimates are usually derived from the knowledge of the radius by means of an approximate version
of the expansive blast-wave evolution equations. We shall see in next Section
that those approximations which ignore the previous phase(s) can introduce 
an apreciable error in cases near phase transition, and such approximations
imply a reduction of the number of variables to two by linking the explosion
energy, the local interstellar medium density and the ejected mass. The present
local density of the interstellar medium can be
estimated within 50\% uncertainty, though this may not be representative of the
structure of the interstellar medium prior to the supernova explosion. for
example, an explosion which occurred inside a HII region, (low density medium) and have run into a nearby molecular
cloud (high density medium) would be misleading, since the observer would be able to estimate the 
molecular cloud density only. Even inside the HII region it would be likely that density
is not constant. Situations like those can be modelled (see
a good review on this in the work of \citealt{TM}), but the point is that the unknown structure
of the pre-supernova environment introduces one more possible systematic error on age 
estimates. The ejected mass can be estimated from the observation of how much Fe was 
produced, coupled to models of supernova yields, when the SNR is young. Explosion
energy is usually considered to be within a factor of 2 of the canonical value of 
10$^{51}$ erg, due to previous both observational and theoretical knowledge. The
review of \citet{SN} considers the energy range $0.5 - 5.5 \; \times 10^{51}$ erg for
classical SNII, from a sample with a dozen well-observed events.
However, these estimates also contribute with their own uncertainties, because
the type of explosion event giving rise to a SGR-AXP is completely unknown. As the
radius picks the same relative uncertainty as the distance, the combined
uncertainty on the age estimate will not be smaller than a factor of 2, and probably
much larger than that. We must recall that it is not uncommon to be unsure about
the phase a given SNR is in,
and that near the transition time between phases approximations do not strictly hold. More accurate
relations can be used, reducing systematic errors, though usually most authors
feel the rather small gain in precision is not worth the trouble.

Blast-wave velocity estimates could also be used to determinate age and radius,
through relations similar to the ones for radius and age. Knowledge of both
radius and velocity would determine the phase of the expansion, removing a source
of uncertainties. Unfortunately, such estimates must come from X-rays measurements
difficult to be made, and up to now they could not be obtained for any of the proposed
associations.

Finally there are evidences of interaction between the neutron star and the SNR in some 
of the proposed associations, which show plerions (or filled-centre morphology).
Additionally, one of the associations was reported as presenting a jet-like
feature \citep{sgr}. 

The overall reliability of the described estimates is not good, although several 
papers have addressed the association issue, and drawn conclusions about AXP and 
SGR nature. In this work, we will show that at least an important factor
for those estimates is still missing, namely the injection of energy by the internal
magnetar, and we analyze the proposed associations
bearing that factor in mind. The next Section is dedicated to SNR expansion
considerations, Section 3 introduces the effects of the injected energy on the SNR, 
while Section 4 deals with an application of those considerations
to the observed sample. In Section 5 we discuss our results.

\section{Supernova Remnant Expansion}

Supernova explosions eject several solar masses to the interstellar medium, which 
form the SNR. In the further evolution, mass from the swept interstellar medium will 
be added, at a rate $4 \pi R^2 \dot R \rho$, where $R$ is the SNR 
radius (blast-wave front), $\dot R$ is the SNR expansion velocity, and $\rho$ 
represents the local interstellar medium density, which we will take as a 
constant for simplicity. A treatment including density as a power-law function 
of the distance from the supernova can be seen on \citet{TM}, which is also a 
good review reference about the non-radiative expansion phases of SNRs.
The initial expansion velocity is set by the total kinetic energy of the explosion,
$E$, and the ejected mass, $M_{ej}$. The interstellar medium will not affect much 
the expansion until the swept mass becomes $\sim \, M_{ej}$, or, equivalently, the SNR mass

\begin{equation}
M \, \equiv \, M_{ej} \, + \, {{4 \pi} \over 3}R^3 \rho \, \sim \, 2 M_{ej} \; .
\label{mass}
\end{equation}

Another approximation that holds true is that very little energy (compared to
the huge kinetic energy of the SNR) is lost, so energy can be considered
constant. In this way, with mass and energy set to constants, and negligible external pressure,
velocity is also (approximately) constant. This is the {\it free expansion}
phase. An approximate linear relation between radius and age ($t$) follows

\begin{equation}
R\, \simeq \, \dot R t \, \simeq \, {\left( {{2E} \over M} \right)}^{1
\over 2} t \; .
\end{equation}

A more accurate expression can be found in the work of \citet{TM}:

\begin{equation}
R\, \simeq \, 0.46 \, {\rm pc} \; E_{51}^{1/2} M_{10}^{-1/2} t_{2} 
{\left( 1+ 0.011 \, E_{51}^{3/4} M_{10}^{-5/4} t_{2}^{3/2} n_1^{1/2} \right)}^{-2/3} \; ,
\label{reltm}
\end{equation}

\noindent
where $n \, = \, \rho m_H^{-1} \mu^{-1}$ is the number density of the 
interstellar medium, $m_H$ is the hydrogen mass and $\mu = 1.4$ is the mean molecular
weight of the interstellar gas. Energy is scaled in units of $10^{51}$ erg
($E_{51} \, = \, E/10^{51}$ erg), mass in units of 10 $M_{\sun}$ ($M_{10} \, = \,
M/10$ $M_{\sun}$), number density in units of 1 cm$^{-3}$ ($n_1 \, = \, n/1$
cm$^{-3}$), and age in units of 100 years ($t_{2} \, = \, t/10^2$ yr).
The velocity at which the blast-wave front expands is approximately

\begin{equation}
\dot R\, \simeq \, 4500 \, {\rm km \; s^{-1}} \; E_{51}^{1/2} M_{10}^{-1/2} 
{\left( 1+ 0.011 \,E_{51}^{3/4} M_{10}^{-5/4} t_{2}^{3/2} n_{1}^{1/2} \right)}^{-5/3} \; .
\label{veltm}
\end{equation}

As $M$ increases, so does the ram pressure of the interstellar medium, slowing
down the SNR expansion. Growing mass and pressure must then be taken into
account. The SNR enters the {\it Sedov-Taylor} phase, named after the works of
\citet{sedov} and \citet{taylor} about pressure-driven explosions. To describe 
the SNR expansion one should solve the equation

\begin{equation}
{d \over {dt}}{\left( {3 \over 4} M \dot R \right)} \, = \, 4 \pi R^2 p \; ,
\label{mov}
\end{equation}

\noindent
the internal pressure being

\begin{equation}
p \, = \, {\left( \gamma -1 \right)} U {\left( {3 \over {4 \pi R^3}} \right)} \; ,
\label{pres}
\end{equation}

\noindent
where $\gamma$ is the adiabatic index (5/3 for an ideal gas, 4/3 for a
relativistic one) and $U$ is the internal energy of the gas inside the internal 
cavity formed by the remnant, whose mass concentrates in a thin shell (its thickness
will be neglected). As the total energy is roughly constant, one can use

\begin{equation}
U \, = \, E \, - \, {9 \over 32} M \dot R^2 \; .
\label{ener}
\end{equation}

Remembering that eq. \ref{mass} shows $M$ depends on $M_{ej}$ and $R$, eq.
(\ref{mov}) turns to have no analytic solution. Nonetheless, as at that phase $M_{ej} \ll M$, 
one could use those approximations to obtain the well-known analytic solution 

\begin{equation} 
R \, \simeq \, {\left( \xi {E \over n} \right)}^{1/5} t^{2/5} \; ,
\label{rela}
\end{equation}

\noindent
where $\xi=2.02$ in the exact solution, and we found 1.77 (both for $\gamma
=5/3$; if $\gamma=4/3$, our solution would be 1.06). 
By construction, this solution disregarded the previous phase. \citet{TM} have calculated a
corrected (though still approximate) expression, which is

\begin{equation}
R \, \simeq \, 12.5 \, {\rm pc} \; M_{10}^{1/3} n_{1}^{-1/3} 
{\left[ E_{51}^{1/2} M_{10}^{-5/6} n_{1}^{1/3} t_{4} -0.051 \right]}^{2/5} \; .
\label{rsttm}
\end{equation}

\noindent
where age was given in units of $10^4$ years ($t_{4} \, = \, t/10^4$ yr). 
The correspondent expression for velocity is

\begin{equation}
\dot R \, \simeq \, 490 \, {\rm km \; s^{-1}} \; E_{51}^{1/2} M_{10}^{-1/2} 
{\left[ E_{51}^{1/2} M_{10}^{-5/6} n_{1}^{1/3} t_{4} -0.051 \right]}^{-3/5} \; .
\label{vsttm}
\end{equation}

Although we have been considering the constancy of the total energy, in fact the
SNR has been slowly radiating away. As the SNR expands, its temperature decreases,
because it depends on the blast-wave velocity, following the well-known relation
for strong shocks,

\begin{equation}
T \, = \, {3 \over 16} {\rho \over {k_B n}} \dot R^2 
\, \simeq \, {3 \over 100} {\rho \over {k_B n}} 
{\left( 2.02 {E \over n} \right)}^{2/5} t^{-6/5} \; ,
\end{equation}
\label{T}

\noindent
where $k_B$ is Boltzmann's constant, and $\dot R$ was taken from the derivative of eq. 
(\ref{rela}).

Eventually T will reach $10^7$ K, where the dominating cooling
process changes from thermal {\it bremsstrahlung} to line emission, which is
more efficient to radiate away energy. That way, the adiabatic approximation ceases
to be accurate, and the SNR enters into the {\it snowplow} phase. As a rough
approximation to estimate when the Sedov-Taylor phase ends, it is usual to estimate how much
time it would take to reduce the thermal energy of the SNR to zero, considering
just radiative losses. Starting from the energy loss per particle,

\begin{equation}
{d \over {dt}} {\left( {3 \over 2} k_B T \right)} \, = \, -n \Lambda \; ,
\end{equation}
\label{rad}

\noindent
where $\Lambda \, = \, 1.6 \times 10^{-19} \zeta T^{-1/2}$ erg cm$^3$ s$^{-1}$ is a
simple cooling function appropriate for ionized gas at temperatures $10^7$ K $\la \, T \,
\la$ 10$^5$ K, $\zeta =1$ is a metallicity factor parametrized for solar 
abundances, one can integrate that equation and equate the result to the dynamical time
$R/\dot R \, \simeq \, 5t/2$, obtaining

\begin{equation}
t_{ST-SP} \, \simeq \, 17 \times 10^3 \, {\rm yr} \; E_{51}^{3/14} n_{1}^{-4/7} \; .
\label{tstspa}
\end{equation}

If we choose to use eq. (\ref{vsttm}) into eq. (\ref{T}), a slightly more
accurate expression results

\begin{equation}
t_{ST-SP} \, \simeq \, 19 \times 10^3 \, {\rm yr} \; E_{51}^{3/14} n_{1}^{-4/7} 
+ 510 \, yr \; E_{51}^{-1/2} M_{10}^{5/6} n_{1}^{-1/3} \; .
\label{tstsp}
\end{equation}

  It has been suggested that SNRs
are very difficult to be seen after 20 kyr \citep{bgl}. The similarity of
the values of ``fading time'' and the transition from the Sedov-Taylor to snowplow
phase may be taken as an indicative that they are related by a factor, of order
unity. Thus, we will adopt eq. (\ref{tstsp}) as the rough limit of visibility of
one SNR. Since our intention is to address associations between SNRs and magnetars, we will not 
explore further the expansion of SNRs into the snowplow phase, because no SNR would
be detectable in that phase or afterwards.

\section{Energy injection by a Magnetar}

The energy loss of a pulsar is usually taken as arising from a rotating magnetic 
dipole approximation,

\begin{equation} 
L_{psr} \, = \, {{B^2 R_{psr}^6 \sin^2 \chi} \over {6 c^3}} 
{\left( {{2 \pi} \over P} \right)}^4 \, = \, 4 \pi^2 I {{\dot P} \over P^3} \; ,
\label{Lpsr}
\end{equation}

That energy is intercepted by the SNR, which will contain, in an internal cavity,
most of either relativistic particles or electromagnetic waves
emmited by the central object. Either way, this cavity would thus be filled by a
relativistic gas, pushing the SNR from the inside.
Considering that the magnetic field, moment of inertia and $\chi$ are constants,
they can be absorbed together with other factors into a new constant 
$K = \, {{B^2 R_{psr}^6 \sin^2 \chi} \over {6 I c^3}}$. After
integrating eq. (\ref{Lpsr}) the period evolution can be expressed as

\begin{equation}
P \, = \, \sqrt{8 \pi^2 Kt+P_0^2} \; ,
\label{P}
\end{equation}

\noindent
with $P_0$ being the initial period of the pulsar. Two additional constants will
be defined

\begin{equation}
\tau_0 \, = \, {P_0^2 \over {8 \pi^2 K}} \, 
= \, 0.6 \, {\rm days} \; 
{\left( {B \over {10^{14} \, {\rm G}}} \right)}^{-2} 
{\left( {P_0 \over {1 \, {\rm ms}}} \right)}^2 \; ,
\label{tau0}
\end{equation}

\noindent
is the initial time-scale for deceleration, and 

\begin{equation}
L_0 = KI{\left( {{2 \pi} \over P_0} \right)}^4 \,
= \, 3.85 \times 10^{47} \, {\rm erg \; s^{-1}} \; 
{\left( {B \over {10^{14} \, {\rm G}}} \right)}^2
{\left( {P_0 \over {1 \, {\rm ms}}} \right)}^{-4} \; ,
\label{L0}
\end{equation}

\noindent
is the initial rate of energy loss. We notice that $L_0 \tau_0^2 \propto B^{-2}$ 
and $L_0 \tau_0 \propto P_0^{-2}$. With these relations and eq. (\ref{P}) we can 
rewrite eq. (\ref{Lpsr}) as

\begin{equation} 
L_{psr} \, = \, L_0 {\left( 1+{t \over \tau_0} \right)}^{-2} \; .
\label{Lmag}
\end{equation}

Total injected energy is just the integral of eq. (\ref{Lmag}) from the initial
instant $t_0 \la 1$ s $< \tau_0$ to the present time,

\begin{eqnarray} 
E_{inj} (t) & = & {L_0 \over {t^{-1} + \tau_0^{-1}}} \nonumber \\
& = & 2.0 \times 10^{52} \, {\rm erg} \; {\left( {P_0 \over {1 \, {\rm ms}}} \right)}^{-2} 
{\left[ 1.6 \times 10^{-3} {\left( {B \over {10^{14} \, {\rm G}}} \right)}^{-2} 
{\left( {P_0 \over {1 \, {\rm ms}}} \right)}^2 
{\left( {t \over {1 \, {\rm yr}}} \right)}^{-1} +1 \right]}^{-1} \; .
\label{Einj}
\end{eqnarray}

Eq. (\ref{tau0}) and (\ref{L0}) are already scaled to the typical magnetar parameters. We
remark that unless magnetars are born with very short periods ($\sim$ 1 ms), their
magnetic fields would not be expected to grow enough to cross the critical quantum
boundary $B_c$, according to the model of magnetar formation of \citet{mag}. While pulsars can be born with such short periods,
they are not required to do so. The difference of two orders of magnitude in the
magnetic field strenght between a typical pulsar ($\sim 10^{12}$ G) and a typical 
magnetar ($\sim 10^{14}$ G) means 4 orders of magnitude in both $\tau_0$ and 
$L_0$ values. This implies in turn a dramatically different influence on SNR 
expansion: a magnetar will inject most of its rotational energy
into the internal cavity of the SNR within a day, while a pulsar would take tens
or hundreds of years, depending on its initial period and magnetic field. Also, a magnetar will
inject typically $10^4$ times more energy than a pulsar, and, more remarkably,
that energy is a factor of 10-20 bigger than the kinetic energy of an ordinary
SNR. It is analogous to the suggestion of \citet{GO} about that transfer of
energy being the very cause of a supernova event.

Gravitational radiation losses are usually larger than rotating magnetic dipole
ones for pulsars with very short periods, hence any estimate of the initial period
based on the electromagnetic dipole torque will result in a figure corresponding
roughly to the period of transition between gravitational and magnetic dipole
dominances. In fact, a simple estimate can be obtained equating eq. (\ref{Lpsr})
to the following expression for gravitational wave-carried energy loss, taken
from \citet{shapiro}:

\begin{equation}
L_{grav} \, = \, {32 \over 5} {{GI^2 \epsilon^2} \over c^5} 
{\left( {{2 \pi} \over P} \right)}^6 \; ,
\end{equation}

\noindent
where G is the familiar gravitational constant and $\epsilon$ is the oblateness
of the neutron star. The result is

\begin{equation}
P_{tran} \, = \, 16 \sqrt{{{3G} \over 5}} 
{{I \pi \epsilon} \over {R_{psr}c B \sin \chi}} \,
= \, 0.34 \, {\rm ms} \; {\left( {B \over {10^{14} \, {\rm G}}} \right)}^{-1} 
{\left( {\epsilon \over 10^{-4}} \right)} \; ,
\end{equation}

\noindent
which shows that for a magnetar the large magnetic dipole losses would not be
much affected by gravitational losses. In other words, the initial
period will effectively be the same period that has allowed the magnetic field
intensity to increase above $B_c$, and almost all rotational energy will be 
injected on the SNR.

However, there is room to consider competition for the energy loss dominance
between magnetodipole radiation and gravitational radiation by r-modes. Doing a very
simple analysis, from the work of \citet{rmod}, we would expect r-modes gravitational
waves to be more efficient than magnetodipole radiation to extract rotational
energy from the rapidly rotating magnetar for $P<4$ ms. Though we feel more detailed studies of
both the possible damping of r-modes by the neutron star crust, and the
enhancement of magnetic fields in newly-born neutron stars are in need before we
can address properly this issue.

Earlier works have addressed pulsar energy injection on a SNR
\citep{port,holan}, but they have not explored the case of a magnetar
source. It can be easily seen that the scenario built by \citet{holan}
cannot hold when magnetars are considered, since they have assumed the energy
injected by the pulsar to be smaller than the kinetic energy of the SNR.

The introduction of magnetar-injected energy changes the equations describing
the expansion of the SNR in the following ways: eq. (\ref{pres}) remains the same,
but $\gamma=4/3$ could be used to represent the dominance of the injected energy
over the initial kinetic energy of the SNR; eq. (\ref{ener}) picks up a new term,
becoming

\begin{equation}
U \, = \, E \, - \, {9 \over 32} M \dot R^2 + {L_0 \over {t^{-1} + \tau_0^{-1}}} \; .
\label{ener-mag}
\end{equation}

However, in this simple formulation of the problem, only the blast-wave front is
described, and we have not described the full problem of two gases. The
energy injection is so quick that we consider all the SNR is instantaneously reacting
to it. That is not strictly true, but since the internal shock (the
inner cavity boundary) would be near the external shock (the SNR boundary) in
less than 100 years, that is much before the transition to Sedov-Taylor phase starts, this
simplification will not affect much Sedov-Taylor and posterior phases. We will
only consider $\gamma=5/3$ through this work. While the works of \citet{port} and
\citet{holan} studied the injection of energy by a pulsar into a SNR, we
adopted simplifications that are not well suited to be combined with their
methods and results. However, the general results obtained in this work are not
expected to be affected by these approximations.

We have performed numerical simulations to find solutions for the set of 
equations describing the position and velocity of the blast-wave front, with and 
without including the energy injected by an internal magnetar, the former case to
test the simulation engine. Our results for the case with energy injection are similar 
to the ones in the case without energy injection, though with explosion energy set to the initial 
kinetic energy plus energy injected (from the internal 
magnetar case). Those results are nearly identical because 
of the extremely short time needed for the injected energy to raise above the 
initial kinetic energy. Our numerical results differ only from the
expressions shown on eqs. (\ref{rsttm}) to (\ref{tstsp}) in one point, namely
our values for $R$ and $\dot R$ in the free expansion phase
differ from \citet{TM} as if the energy were reduced by a factor of 1.14. 
This factor comes from the approximation done on Section 2, specifically the
difference in $\xi$ values on eq. (\ref{rela}). 

This factor must be multiplied to the energy if our values are to be
compared to \citet{TM}. Through this work, however, we will adopt the values we
have found, without corrections. The discrepancy is not large, and certainly
does not affect much our considerations, given other simplifications done in this work.

\begin{figure}
\plotone{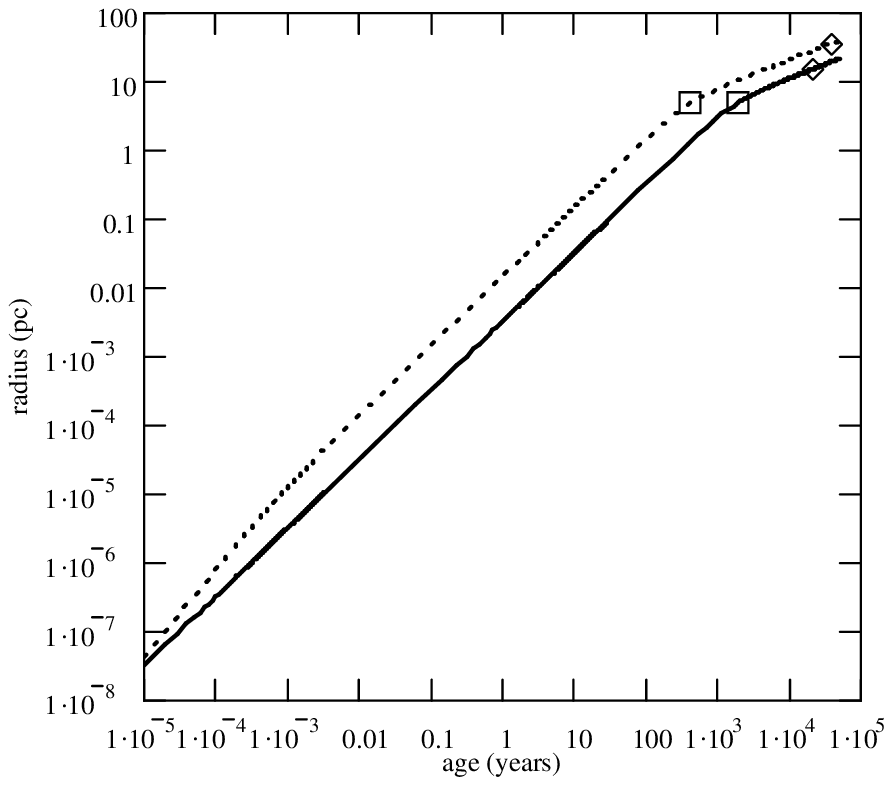}
\caption{Radius evolution for a supernova with $M_{ej}=10$ M$_\sun$, $n=1$
cm$^{-3}$, and $E=10^{51}$ erg. The solid line represents the case without
energy injection, and the dotted line represents the case with energy injection
by a magnetar with $B=5\times 10^{14}$ G and $P_0=1$ ms. Squares mark the
transition to Sedov-Taylor phase, and diamonds mark the transition to snowplow
phase, where these curves are no longer valid. \label{f-rt}}
\end{figure}

\begin{figure}
\plotone{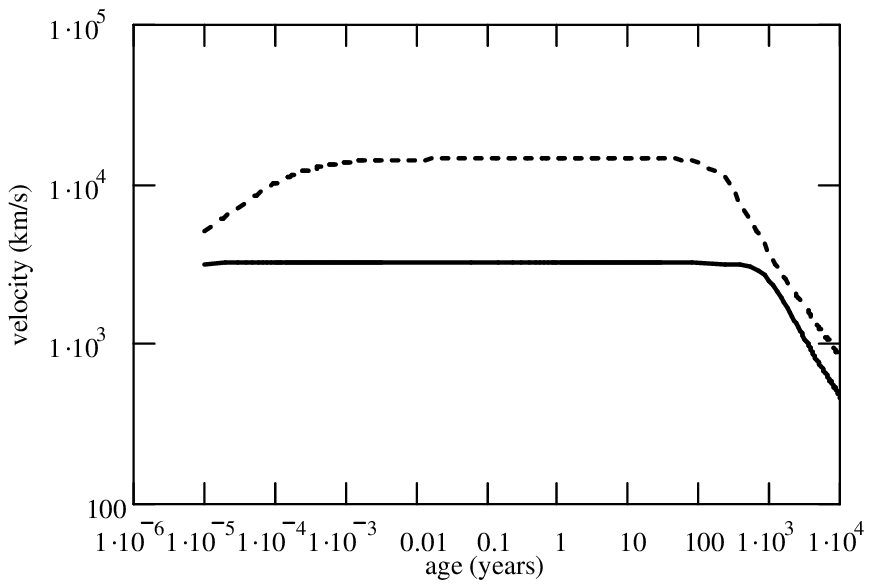}
\caption{Expansion velocity evolution for a supernova with $M_{ej}=10$ M$_\sun$, $n=1$
cm$^{-3}$, and $E=10^{51}$ erg. The solid line represents the case without
energy injection, and the dotted line represents the case with energy injection
by a magnetar with $B=5\times 10^{14}$ G and $P_0=1$ ms. \label{f-vt}}
\end{figure}

The numerical results can be appreciated in Figs. \ref{f-rt} and \ref{f-vt}. In
Fig. \ref{f-rt} we show the evolution of $R(t)$ for both cases (with and
without energy injection). In Fig. \ref{f-vt} we show the evolution of $\dot
R(t)$ for both cases.

\notetoeditor{Figures \ref{f-rt} and \ref{f-vt} could appear side-by-side in print} 

\section{Analysis of Proposed Associations}

\begin{figure}
\plotone{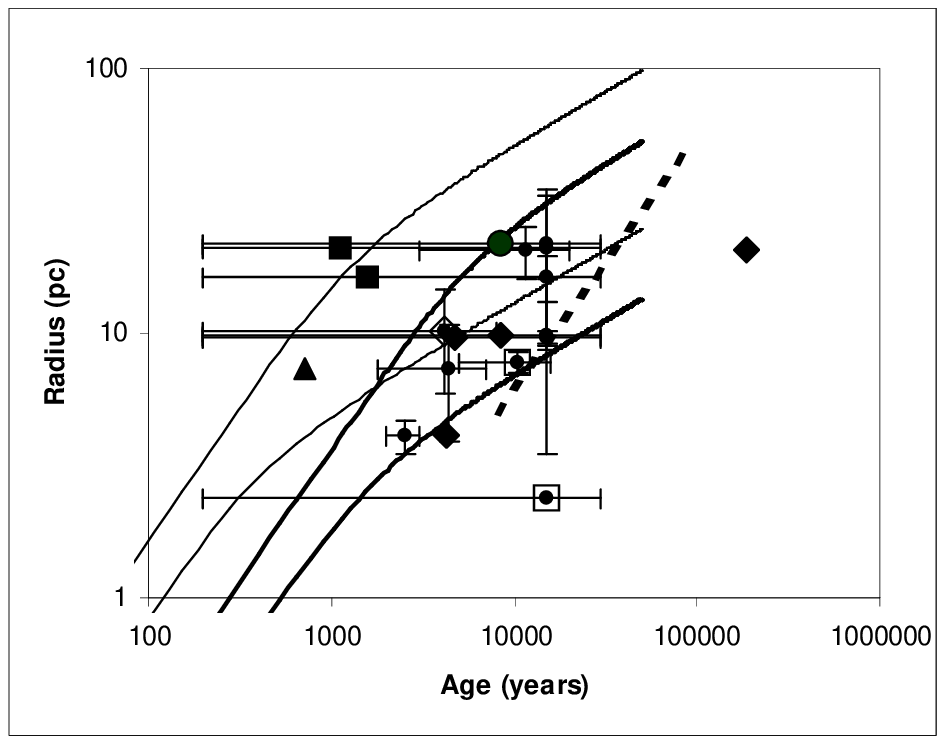}
\caption{Comparison of age and radius ranges for the proposed associations to
the models with and without energy injection. The thick solid lines represent
radius evolution for the two standard extreme cases (``low density/low mass'' and
``high density/high mass'' scenarios,
see text). The dotted thin lines represent the evolution with energy injection
for the two extreme cases. The thick dashed line is the approximate end of
Sedov-Taylor phase for the model without energy injection. SGRs (filled squares)
and AXPs (filled diamonds) with estimated $\dot P$ values are placed according to the median of their
characteristic age ranges and the median of the radius ranges for the associated
SNRs. SGRs (open squares) and AXPs (open diamonds) without estimated ages are
placed on the median values for their associated SNRs. SNRs are small dots
placed on the median values of the ranges
(shown as error bars). For SNRs with unreliable ages, we assumed an
arbitrary range 0.2-30 Kyr. The new association proposed for AXP 1709-4009
is marked as a filled circle. PSR J1846-0258 is marked as a triangle. \label{f-dados}}
\end{figure}

The proposed associations between would-be magnetars and SNRs are shown in Tables
\ref{t-as} and \ref{t-snr}. We shall analyze them by looking first into the situation
without considering the injection of energy by a magnetar. This situation is
shown in Fig. \ref{f-dados}, where we show the range of radius and ages
associated to each SNR considered, the characteristic ages of the magnetars,
and the radius evolution curves for two different cases (thick solid lines):
a ``low density/low mass'' evolution scenario, in which $M_{ej}=8$ M$_{\sun}$ and $n=0.01$ cm$^{-3}$, and 
a ``high density/high mass'' evolution scenario, with $M_{ej}=30$ M$_{\sun}$ and $n=10$
cm$^{-3}$. For both extreme cases, explosion energy is held fixed to 10$^{51}$ erg. 
The position of a given SNR should be between these two extremes, unless quite
different explosion energies are considered.
Certain SNRs have no reliable age estimate \citep{nn}; for those, we adopted an
arbitrary range 0.2-30 kyr, which is the possible range for galactic SNRs.
 For several associations, SNR and magnetar ages
are not compatible, and some magnetar ages are not compatible with the SNR
expansion in enviroments with ordinary values for $n$. That has been used to
justify the model of AXPs and SGRs being born in regions of higher density than
radio pulsars \citep{nn}, or to dismiss the association altogether.

Varying the ejected mass value will displace those curves diagonally
(in free expansion phase only), but will not affect meaningfully the position of the 
curves in the Sedov-Taylor phase. In any case, quite unreasonable values of both $M_{ej}$ 
and $n$ should be invoked to maintain some associations as valid. The other way to
displace the curves (in both phases) is by changing explosion energy.

\notetoeditor{We suggest Figure 3 come here} 

In Fig. \ref{f-dados} we also show the curves 
including the energy injection by a magnetar with $B=5 \times 10^{14}$ G and
$P_0 = 1$ ms. The displacement of curves helps to attribute lower values to the density
than before to all associations, and it makes possible to ``save'' some otherwise
untenable associations.
Thus, the injection of energy could be behind the age discrepancy, as
SNRs truly associated with magnetars would have expanded faster than expected.
In other words, the SNR which has been born with an internal magnetar will seem 
older. The actual relation between the true age ($t_t$) and the apparent
one ($t_a$) can be obtained from eq. (\ref{rsttm}), for the Sedov-Taylor
phase, considered for both the conventional energy value ($E_a$) and the one 
including energy injected ($E_t$), as

\begin{equation}
t_t \, = \, \sqrt{{E_a \over E_t}} t_a \; .
\label{te}
\end{equation}

\noindent
Typical figures would be in the range $t_t \sim 0.2-0.3 \, t_a$.

Likewise, other reference timescales will be shifted. The transition from free
expansion to Sedov-Taylor phases will occur sooner, because of the higher
initial velocity. The transition to snowplow phase will occur at 36 kyr
after the supernova explosion (see eq. \ref{tstsp}), using $E_t = 20 E_a$. This
is just meant to show that an SNR with energy injection could be visible for 
more time than an ordinary SNR, while at the same time appearing to be younger. The time a
neutron star will take to catch up its SNR ($t_{cross}$) before the 
transition to the snowplow phase (while it is still visible) can 
be roughly estimated from eq. (\ref{rsttm}) and the neutron star 
(constant) velocity $v=10^2 v_{2}$ km s$^{-1}$, resulting in

\begin{equation}
t_{cross} \, = \, 6.5 \times 10^5 \, {\rm yr} \; v_2^{-5/3} E_{51}^{1/3} n_{1}^{-1/3} \; 
\label{tcross}
\end{equation}

\noindent
where we ignored the small negative term on eq. (\ref{rsttm}) for the sake of
simplicity. From this last equation it can be seen that only a fast magnetar can catch up 
its SNR shell. For example, if $v_2=10$, as suggested for some 
associations, and $E_{51}= 20$, then $t_{cross} \simeq 38$ kyr. Of 
course, the proximity of the neutron star to the SNR shell can induce 
a ``reenergization'' of the shell, extending the visibility of the SNR to 
later ages (see \citealp{re-snr}). We do not address this possibility here.

Writing the distance of the neutron star to the center of the SNR as $r(t)=\beta (t) R(t)$, 
the quantity $\beta$ becomes an observable parameter (projected in the sky). Inverting 
eq. (\ref{tcross}) and inserting $\beta$, we can find an expression for the minimal neutron 
star velocity needed to reach a relative displacement $\beta$ at age $t_4$,

\begin{equation}
v \, = \, 1.2 \times 10^3 \, {\rm km \; s^{-1}} \; \beta \, t_4^{-3/5} E_{51}^{1/5} n_{1}^{-1/3} \; .
\label{vbeta}
\end{equation}

\noindent
If the neutron star velocity points transversally to the line of sight, then 
eq. (\ref{vbeta}) gives the actual velocity of the neutron star.

\subsection{Analysis of magnetar candidates with known $\dot P$}

We shall analyse the proposed associations taking into account the energy 
injected by the internal magnetar, addressing the plausibility of the 
association, and the nature of the compact object. Similar studies have been 
published by several authors \citep{mer99,nn,gae01}, based mainly on position 
and velocity considerations. Those works have disregarded magnetar 
characteristic ages, in favour of SNR estimated ages. However, their 
conclusions are discrepant. We shall in turn disregard the ages of the SNRs, 
because if they were born with an internal magnetar, their ages are 
overestimated according to eq.(\ref{te}).

The novel feature of our approach is just the fact that, if the AXPs and SGRs are 
magnetars, following the model of \citet{mag} for magnetar formation, which is based
on a dynamo action to make possible the growth of the magnetic field beyond ordinary
pulsars range, they will inject energy enough on their SNRs to affect their 
expansions, making them reach a larger size in less time than it was considered 
by previous authors.
Although our simulation engine is probably too crude to provide a reliable estimate
of the time after which the injection of energy becomes less efficient, even
more sophisticated simulations have not been tested to know this number. We will take
as enough to say that injection times below one month are securely within the bounds to provide
optimal coupling between injected energy and the SNR kinetic energy.
As the main effect of the energy injected is to add to the 
initial kinetic energy of the SNR, any pulsar which can inject most of its 
rotational energy in less than one month will provide an SNR evolution as if the 
kinetic energy were the initial plus the injected one.
The injection time for the smallest $B$ of the AXP/SGR sample (AXP 2259+586) is lower
than 5 days, considering $P_0 \sim 1$ ms. Even initial periods as large as 3 ms 
would provide injection times close or within to the one month figure, which make 
us certain of the effectiveness of the coupling. Pulsars, on the other hand, have
injection times longer than decades, typically.

 For simplicity, and 
having no way to estimate the actual initial periods of AXPs and SGRs, we 
assume all them as being born with $P_0=1$ ms, which in turn determines that the 
injected energy ($\sim 2 \times 10^{52}$ erg at $t >> \tau_0$) will be 
essentially the same, regardless of the actual value of $B$ (from eq. 
\ref{Einj}), as long as $\tau_0 \la$ 1 year. Because of this fact, we will only 
consider two extreme situations to analyze the associations: the ``low density/low
mass'' evolution scenario, where $M_{ej}=8$ M$_{\sun}$ and $n=0.01$ cm$^{-3}$, and 
the ``high density/high mass'' evolution scenario, with $M_{ej}=30$ M$_{\sun}$ and $n=10$ cm$^{-3}$. 
The actual situation for each case should be bracketed between these values. The 
result can be seen in Fig. \ref{f-known}, in which the range of radius and 
characteristic ages values for each association is explicitely shown. Our
assesment of each association is as follows:

{\bf SGR 1806-20/G10.0-0.3}: The probability of alignment by chance is $\sim 
0.5\%$ \citep{nn}. G10.0-0.3 was considered as not being a SNR 
\citep{gae01,chak}, although \citet{green} lists it in his catalogue. A cluster of
stars is close to the line of sight to this association \citep{fuc}, so either the
SGR or the SNR may be physically related to it. In our 
model, considering the characteristic age of SGR 1806-20 as the true age of the 
association, it can be seen that the entire range of radius values lie 
between the two extreme scenarios (Fig. \ref{f-known}), with an 
indication of mid to low-density interstellar medium. The magnetar transversal 
velocity implied is high, 4000-6500 km s$^{-1}$, if $\beta=0.5$ \citep{k94}. However, if 
$\beta \sim0$ \citep{chak}, the velocity cannot be inferred. G10.0-0.3 would be 
entering Sedov-Taylor phase. Dropping altogether the characteristic age 
as a good age estimator, the association could be as old as 15 kyr, with 
$v\simeq 500$ km s$^{-1}$ ($\beta=0.5$). This association can be considered as 
true if SGR 1806-20 is a magnetar, $\beta \sim0$, and G10.0-0.3 is confirmed as
a SNR.

{\bf SGR 1900+14/G42.8+0.6}: The probability of random alignment is $\sim 4\%$ 
\citep{gae01}. A pulsar was recently discovered near that position \citep{lx}, 
that could be related to the SNR, although its characteristic age is 38 kyr. 
Our model would allow for the association if the true distance were on the 
lower 1/3 of the range and the true age were on the upper 1/2 of the quoted range, 
but the extremely high velocity implied ($\geq 8000$ km s$^{-1}$) precludes 
that. Disregarding characteristic age, the range would be extended up to 6-40 kyr, 
depending on true distance and interstellar medium density, though the magnetar 
velocity would still be high (800-2000 km s$^{-1}$). \citet{china} suggested
that SGR 1900+14 was born in the 4 B. C. supernova, and the discrepancy between 
characteristic age and the proposed age of 2 kyr is attributted to dynamical 
evolution with braking index $\simeq$2, but they did not offer a good reason
to explain how the SNR disappeared from view in just 2 kyr. We conclude that this 
association is not convincing (if SGR 1900+14 is a magnetar), unless a mechanism
for a high velocity of the magnetar is adopted.

{\bf AXP 1048-5937/G287.8-0.5}: The probability of chance alignment is 
$\sim 16\%$ \citep{nn, gae01}. Data about G287.8-0.5 was considered unreliable 
by \citet{gae01}. Nevertheless, if the SNR is confirmed as such, our model indicates that it 
is entering the Sedov-Taylor phase, on a high density interstellar medium. 
Low ages are preferred. Again, the very high velocities implied (4500-8500 km 
s$^{-1}$) argue against the association. In this case, even to disregard the 
characteristic age does not improve the plausibility of the association. We 
conclude that this association is unlikely.

{\bf AXP J1709-4009/G346.6-0.2}: The random alignment probability is $\sim 
10\%$ \citep{nn} to $\sim 30\%$ \citep{gae01}. Our model could allow the 
association if the interstellar medium has high density, and the true distance 
is on the upper 25\% values of the range. Once more, the high velocity 
required for $\beta=1.7$ (3000-2000 km s$^{-1}$) is in excess of the known 
pulsar population. Disregarding the characteristic age would only help very 
slightly to avoid the ``velocity problem''. We therefore consider this 
association as unlikely.

{\it AXP J1709-4009/G346.5-0.1}: \citet{gae01} suggest this association 
instead of the previous one. While the newly-identified SNR G346.5-0.1 awaits 
to be confirmed, we can analyze the association in the same fashion we have 
been doing. The probability of random alignment is $\sim 10\%$ \citep{gae01}. 
Our model allow the association if the true distance is on the upper 80\% 
values of the range. Once more, the high velocity required for $\beta=1.2$ 
(1700-4800 km s$^{-1}$) exceeds the typical pulsar velocity. Disregarding the 
characteristic age would allow the age range to go up to $\sim50$ kyr 
($\sim800$ km s$^{-1}$), at the transition to snowplow phase. We 
consider this association as more likely, requiring a mid to high-density medium. 
We notice that our solutions with injection are not better than the old situation (without 
energy injection), so the association can be true even if AXP J1709-4009 is 
not a magnetar.

{\bf AXP 1841-045/G27.4+0.0}: The chance alignment probability is $\simeq 0.01\%$ 
\citep{gae01}. In our model, this association would require an excedingly low 
density interstellar medium, or negligible energy injection. Equivalently the
association would require $P_0>6$ ms, or age$\leq$ 600 yr (but with $v\geq$ 1600 km s$^{-1}$). 
Even then, the characteristic age would have to be disregarded. Contrary to 
most AXPs and SGRs, the characteristic age of AXP 1841-045 is higher than the 
SNR estimated age. The association is possible only if 
AXP 1841-045 is not a magnetar with short $P_0$.

{\bf AXP 2259+586/CTB 109}: The probability of random alignment is $\sim 0.05\%$ 
\citep{gae01}. The characteristic age of AXP 2259+586 puts its associated SNR 
in the snowplow phase, where it would be unlikely to be detected. Given 
the low probability of chance alignment, this age discrepancy (this is the 
other AXP which has characteristic age higher than SNR age) argues again against the 
characteristic age as a good estimate of the true age. Removing this parameter 
allows for our model a wide range of possible age values, from 1 to 30 kyr. The 
requirement of reasonable magnetar velocity would limit ages $\leq6$ kyr.
Moreover, this range is already allowed by the expansion without energy injection. The 
association is probably true for both models.

\subsection{Analysis of magnetar candidates with unknown $\dot P$}

For these objects, we have no information about their characteristic ages or 
magnetic field strenght. Therefore the task here is to verify if the allowed range of ages of 
our model is compatible with reasonable magnetar velocities.

{\bf SGR 0526-66/N 49}: The random alignment probability is $\sim 0.7\%$ \citep{gae01}. 
The allowed range of ages is 0.5-3.5 kyr, which implies extremely high velocities, 
$v>1800$ km s$^{-1}$. If this association is true, then SGR 0526-66 would have 
to be another magnetar born with a long $P_0$, as suggested for AXP 1841-045.

{\bf AXP J1845-0258/G29.6+0.1}: The chance alignment probability is $\sim 0.2\%$ \citep{gae01}. 
The allowed range of ages is 1-13 kyr, which implies on $3500>v>300$ km s$^{-1}$,
for the largest distance of the range, and 0.4-1.5 kyr 
($3600>v>1000$ km s$^{-1}$) for the opposite extreme of the distance range. 
This association can be true with or without energy injection. If AXP 
J1845-0258 is a magnetar, then G29.9+0.1 prefers the higher end of values for 
distance and age.

{\bf SGR 1627-41/G337.0-0.1}: The probability of random alignment is $\sim 5\%$ 
\citep{gae01}. The allowed age range of our model to this association (150-300 
yr) imply on $v>15000$ km s$^{-1}$, ruling out the magnetar hypothesis, unless 
$P_0\geq 6$ ms, as discussed for AXP 1841-045. Even so, the velocity problem 
still holds. Thus, we find the association unlikely.

{\bf SGR 1801-23/G6.4-0.1}: SGR 1801-23 is just a candidate SGR, and its 
position is not well-determined. The distance to G6.4-0.1 is uncertain, too. 
Within such set of data, our model allows for wide ranges of ages. In the high extreme 
of the distance range, ages can be 2-35 kyr ($1000>v>70$ km s$^{-1}$). In the 
low end of the distance range, ages allowed are 0.45-2 kyr ($1500>v>350$ km 
s$^{-1}$). Thus, the association is probable, and this holds even if SGR 1801-23 is not a 
magnetar at all. Though, this result is heavily dependent on a better position
determination.

\section{Discussion}

We have analyzed in this work the proposed SGR-AXP/SNR associations. Previous
analisis of the associations arrived at different results. According to 
\citet{nn}, all associations can be considered likely. \citet{gae01} contend
that only AXP J1845-0258/G29.6+0.1, AXP 1841-045/G27.4+0.0 and AXP 2259+586/CTB 109 could 
be valid. \citet{ank} considered SGR 1806-20/G10.0-0.3 and SGR 0526-66/N 
49 as plausible, in addition to the 3 already mentioned by \citet{gae01}.

It is expected that if magnetars exist and are born in supernova explosions,
they can inject enough energy to enhance the expansion of their associated SNRs. If 
AXPs and SGRs are indeed magnetars, their associated SNRs should appear older
than the ages derived from standard expansion models. We analyzed all the proposed 
associations, and have come to the following results:

1) If the characteristic age of the neutron star is regarded as the true 
age of the association, then SGRs and AXPs may not be magnetars at all (or the model
of \citet{mag} is not correct regarding magnetar origin), for only one case 
(SGR 1806-20/G10.0-0.3) has shown good agreement within the model, and even the 
true nature of G10.0-0.3 was put in doubt;

2) If characteristic ages are in turn ignored (see 
\citealt{venrel,gae01,nn,kes75}, among others), then two SGRs (out of 5) 
and three AXPs (out of 5 with proposed associations) seem to be associated
within our model, so about 50\% of general agreement. It should be noted that
two associations which agreed within our model (AXP J1845-0258/G29.6+0.1 and AXP 
2259+586/CTB 109) were already believed to be true by previous works 
\citep{gae01,nn,ank}, meaning that uncertainties in distance are large enough 
to allow for both possible scenarios (standard and with energy injection);

3) AXP 1841-045/G27.4+0.0 and SGR 0526-66/N 49 can only be considered as true 
associations if the magnetars were born with 
$P_0 \geq 6$ ms, because in that way the energy injected would be insufficient 
to directly affect SNR expansion.

These results from our model are tied to the dynamical evolution of magnetars. 
While we have assumed for simplicity the standard magnetic dipole braking with 
braking index equal to 3, several proposals have been done that argue for different 
braking models: fossil or fallback accretion disks \citep{nn,CHN00,PTH95}, 
episodes of relativistic wind emission \citep{venrel}, a different constant 
braking index \citep{china}, magnetic field decrease \citep{mag,mon}. As the 
measurements of braking indices for 5 young pulsars revealed that all but one 
have significant departures from the canonical value, it should not come as a 
surprise that magnetars do not follow the standard model of spindown. The 
main effects of alternative models are to change the estimates of magnetic 
field strength and spindown age. Nonetheless, the influence in our model 
would be very small, since $\tau_0$ can be increased up to one month 
without appreciable modifications on the model, and rotational energy is not dependent on 
spindown models. That is why we feel justified to leave item (1) above aside of 
the discussion. Spindown models that allow for ages significantly smaller or larger
than the conventional characteristic age are required to explain certain
associations, if they are true ones, either within our model, or considering the
standard scenario.

Supposing that our results from item (2) above represent the actual situation, 
we could ask where are the SNR that are associated to the other SGRs and AXPs. 
The answer can be that they were not detected yet, since they lie generally at 
regions of coexistence of HII regions, other SNRs, and other types of objects 
(variable stars, young stars, molecular clouds, etc.). Including AXP 0142+614 
and AXP J0720-312 in the sample, we would have slightly less than 50\% of the true 
associations identified. On the other hand, few of the latter can be 
considered as firm, due to several other factors. For example, the 
angular size of SNRs are often quoted without uncertainties, which certainly is 
an understatement, given the dificulty to recognize a SNR and to assess its shape and 
size.

It is worthy to remark that a similar analysis, ignoring the effects of the injection
of energy, and desconsidering conventional characteristic ages, can be done to the
same associations, and will result in different age and velocity ranges than those
obtained here. Indeed, even associations considered unlikely by us may turn out to be
likely. This is because the standard energy SNRs would take more time to reach the
observed sizes, allowing for higher ages for the associations than in our model, and
thus producing lower velocities than those we found. Nevertheless, that scenario
requires non-standard spindown and no injection of energy by the neutron star, be it
a magnetar or not.

It is interesting to check that the most important factor to decide on the
plausibility of associations is $\beta$. As
$\beta$ increases, associations are considered increasingly unlikely. However,
\citet{gvar} points to the possibility that a SNR expanding in a region with
anisotropic interstellar medium densities will be distorted and/or expands
faster in one or more directions. This means that the geometrical center of the
SNR can be displaced from the actual explosion site. Although it is difficult to
take this effect into account quantitatively, one should be cautious to do not dismiss it
entirely. It is possible that one or more of the proposed associations are
affected by this effect, which can increase or decrease $\beta$ randomly. That effect will
be more accentuated for SNRs that have been expanding in low-density regions
(``bubbles'') amidst high-density ``walls'', and it increases with age. The
first consideration is the case for the most massive stars, which do not live
long enough to move far away from the sites where they had been born, while their
stellars winds create a low-density cavity. The second consideration means we
must expect $\beta$ estimates to be a little bit more scattered as age (and
radius) increases.

\begin{figure}
\plotone{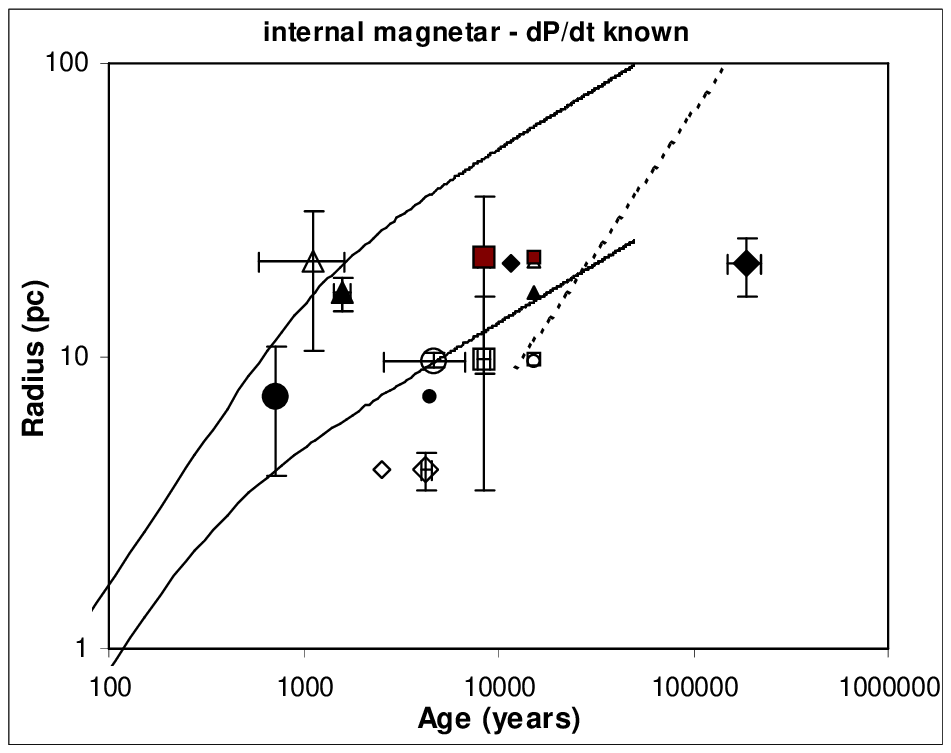}
\caption{Comparison of radius and ages of the proposed associations to the
expansion model with energy injection, for objects with estimated $\dot P$. The solid 
lines represent the evolution according to our model. Large markers 
are placed according to the median of characteristic ages and radii ranges. Small markers 
are placed according to the median of estimated SNR ages and radii. Error bars
indicate ranges of characteristic ages and SNR radii. The dotted line is the
approximate end of Sedov-Taylor phase (including energy injection). Filled
markers represent likely associations, and open markers represent the unlikely
ones (see text). The markera are: open triangle (SGR 1900+14), filled triangle (SGR 1806-20), 
open circle (AXP 1048-5937), open square (AXP 1709-4009), open diamond (AXP 1841-045),
filled diamond (AXP 2259+586), filled square (AXP 1709-4009, new association),
filled circle (PSR J1846-0258). \label{f-known}}
\end{figure}

Besides studying associations including AXPs and SGRs, we also examined the
association PSR J1846-0258/G29.7-0.3 \citep{gvk75,kes75}, which also suffers from the age 
problem, since it has the smallest characteristic age known among pulsars (723 
years), and the SNR age was estimated as at least 1800 years. The magnetic field of this 
pulsar is $\sim 5 \times 10^{13}$ G, considering magnetic braking spindown, slightly 
above the quantum critical field, and by so marginally qualifying as a magnetar. 
It was not detected at radio wavelenghts, only in X-rays. The association is 
considered very likely, as the pulsar is located at the geometrical center of the 
SNR, coincident with a radio/X-ray nebula, probably powered by the pulsar. Proceeding 
as in the other cases investigated in the previous Sections, we see from Fig.
\ref{f-known} that the range of radius quoted on the literature is nearly
coincident with the range allowed by our two extreme cases, considering the
characteristic age. This way, there are no
preferences for high or low densities or ejected mass. As before, we will ignore
the characteristic age, to found that the age range allowed is 250-700 years, in
case of high $M_{ej}$ or $n$, and 650-6000 years, in case of low $M_{ej}$ or
$n$. Intermediate values of the parameters $M_{ej}$ and $n$ would provide
intermediate ranges of ages. The placement of the neutron star at the
geometrical center of the SNR imply in low velocities or alignment between the
velocity vector and the line of sight. It is important to notice that \citet{kes75}
find a braking index $\sim 1,9$ and age $\sim 1700$ years for this pulsar, in
which case the magnetic field can be under the quantum critical one. Nonetheless,
this association represents one more evidence against the consideration of the 
characteristic age (without braking index information) as a good age estimate.

We have left for a future work (Allen \& Horvath, in preparation) the study of an
alternative origin scenario for magnetars, the collapse of a white dwarf star, 
induced by accretion or merging from a binary companion. Simulations for this scenario
\citep{aic} reveal that $\sim 0.1$ $M_{\sun}$ can be ejected, with an explosion
energy of $10^{50}$ erg, implying on higher initial velocities.

Finally we would like to point that magnetars lose very little rotational energy through
gravitational waves, when compared to typical pulsars, unless r-modes can play
an important role, which is not clear to us. Statistical studies on pulsar gravitational 
wave detectability, like the one performed by \citet{pach}, shall not be affected
by this consideration, at least for the current detectors, like VIRGO, because
of the scarcity of magnetars.

\acknowledgments

We are grateful to G. Medina-Tanco for discussions about SNR blast-wave physics. This
work was supported by FAPESP Agency (S\~ao Paulo State, Brazil) and CNPq (Brazil).

\clearpage

\begin{deluxetable}{lccccc}
\tablewidth{0pt}
\tablecaption{Data of SGRs and AXPs \label{t-as}}
\tablehead{ 
\colhead{magnetar} & \colhead{$P$ (s)} & \colhead{$\dot P$ ($10^{-12}$)} &
\colhead{$\tau$ (kyr)} & \colhead{$B$ ($10^{14}$ G)} & \colhead{$d$ (kpc)} }
\startdata
\sidehead{SGRs}
SGR 1806-20 & 7.48 $^i$ & 83 $^i$ & 1.4 $^i$ & 8.0 $^i$ & 17 $^c$, 14 $^l$ \\
SGR 1900+14 & 5.16 $^j$ & 50 - 140 $^q$ & 0.58 - 1.6 $^q$ & 5.1 - 8.6 $^q$ & 7 $^j$, 5 $^k$ \\
SGR 0526-66 & 8.00 $^a$ & \nodata & \nodata & \nodata & 55 $^k$ \\
SGR 1627-41 & 6.4 ? $^{B}$ & \nodata & \nodata & \nodata & 11 $^A$ \\
SGR 1801-23 & \nodata & \nodata & \nodata & \nodata & \nodata \\
\sidehead{AXPs}
AXP 1048-5937 & 6.45 $^m$ & 15 - 40 $^q$ & 2.6 - 6.8 $^q$ & 5.1 - 8.6 $^q$ & 10.6 $^n$, $>2.8$ $^q$ \\
AXP J1709-4009 & 11 $^{qt}$ & 19 $^{t}$, 22.5 $^v$ & 9.2 $^{t}$, 7.7 $^{v}$ & 4.6 $^{t}$, 5.0 $^{v}$ & 10 $^m$, $>8$ $^q$ \\
AXP 1841-045 & 11.8 $^d$ & 47 $^h$, 41 $^{d}$ & 3.9 $^h$, 4.5 $^d$ & 7.6
$^h$, 7.0 $^d$ & 7 $^{h}$ \\
AXP 2259+586 & 6.98 $^{n,t}$ & 0.74 $^n$, 0.49 $^t$ & 150 $^n$, 230 $^t$ & 0.73 $^n$, 0.59 $^t$ & 6.2 $^n$ \\
AXP J1845-0258 & 6.97 $^E$ & \nodata & \nodata & \nodata & 8.5 $^E$\\
\sidehead{other}
PSR J1846-0258 & 0.324 $^X$ & 7.1 $^{X,Y}$ & 0.72 $^{X,Y}$ & 0.5 $^{X,Y}$ &
\nodata \\
\enddata
\tablerefs{
(a) \citet{mag}; (c) \citet{kou94}; (d) \citet{gv97}; (h) \citet{vg97}; (i)
\citet{sgr}; (j) \citet{kou99}; (k) \citet{hur99}; (m)
\citet{chak}; (n) \citet{PTH95}; (o) \citet{gae00}; (p) \citet{kas00}; (q)
\citet{mer99}; (t) \citet{kaset99}; (v) \citet{is}; (A) \citet{CCDD99}; 
(B) \citet{wo99}; (E) \citet{TKKTY}; (X) \citet{gvk75}; (Y) \citet{kes75}.} 
\end{deluxetable}

\clearpage

\begin{deluxetable}{lcccc}
\tablewidth{0pt}
\tablecaption{Data of associated SNRs \label{t-snr}} 
\tablehead{ 
\colhead{SNR} & \colhead{$t_a$ (kyr)} & \colhead{$d$ (kpc)} & \colhead{$R$ (pc)}
& \colhead{$\beta$} } 
\startdata 
\sidehead{associated to SGRs}
G 10.0-0.3 & 0.2 - 30 $^r$ & 13 - 16 $^D$ & 14 - 19 $^D$ & 0.0 $^m$, 0.6 $^b$ \\
G 42.8+0.6 & 0.2 - 30 $^r$ & 3 - 9 $^r$ & 11 - 31 $^r$ & 
1.2 $^m$, 1.4 $^{r,s}$ \\
N 49 & 5 - 16 $^{r,s}$ & 55 $^f$, 50 $^{g,z}$ & 7.1 - 8 $^r$, 8.5 $^f$ & 0.6 $^m$, 1.0 $^s$ \\
G 337.0-0.1 & 0.2 - 30 $^r$ & 11 $^{e,s,A}$ & 2.4 $^{e,r}$ & 1.7 $^s$, 2.3 $^{r,A}$, \\
G 6.4-0.1 & $>2.4$ $^r$ & 1.2 - 3 $^r$, 3.5 - 4 $^e$ & 7 - 17.5 $^r$, 21 - 24 $^e$& 0.1 $^r$ \\ 
\sidehead{associated to AXPs} 
G 287.8-0.5 & 0.2 - 30 $^r$ & 2.5 - 2.8 $^r$ & 9.1 - 10.2 $^r$ & 2.2 $^r$ \\
G 346.6-0.2 & 0.2 - 30 $^r$ & 3 to 5 $^r$, 11 $^s$ & 4.4 to 7.3 $^r$ & 1.7 $^r$ \\
G 27.4+0.0 & 2 $^s$, $<3$ $^q$ & 6 to 7.5 $^{e,w}$ & 3.5 - 4.4 $^{e,r}$, 4.7 $^{h}$ & $<$0.25 $^s$, 0.1 $^r$, \\
G 109.1-1.0 & 3 to 20 $^q$ & 4 to 5.6 $^{q,y}$ & 16 - 24 $^r$, 18 - 25 $^y$
 & 0.3 $^m$, 0.2 $^r$, $<$0.2 $^s$ \\
G 29.6+0.1 & 0.2 - 30 $^r$ & 8.5 $^m$, $<20$ $^{q,s,x}$ 
& 6.5 to 9.8 $^r$ & 0.1 $^r$, $<0.25$ $^s$ \\
\sidehead{associated to other} 
G 29.7-0.3 & 1.8 - 7 $^Y$ & 9 - 21 $^e$ & 3.9 - 10.7 $^{e,X}$ & 0.0 $^Y$ \\
\enddata
\tablerefs{
(b) \citet{k94}; (d) \citet{gv97}; (e)
\citet{green}; (f) \citet{mat83}; (g) \citet{shu}; (h) \citet{vg97}; 
(j) \citet{kou99}; (k) \citet{hur99}; (m)
\citet{chak}; (o) \citet{gae00}; (q)
\citet{mer99}; (r) \citet{nn}; (s) \citet{gae01}; (w) \citet{SH92}; (x) 
\citet{GGV99}; (y) \citet{RP97}; (z) \citet{VBLR}; (A) \citet{CCDD99}; (D) 
\citet{cor97}; (X) \citet{gvk75}; (Y) \citet{kes75}.}
\end{deluxetable}

\clearpage

\clearpage

\clearpage

\clearpage

\clearpage

\begin{figure}
\plotone{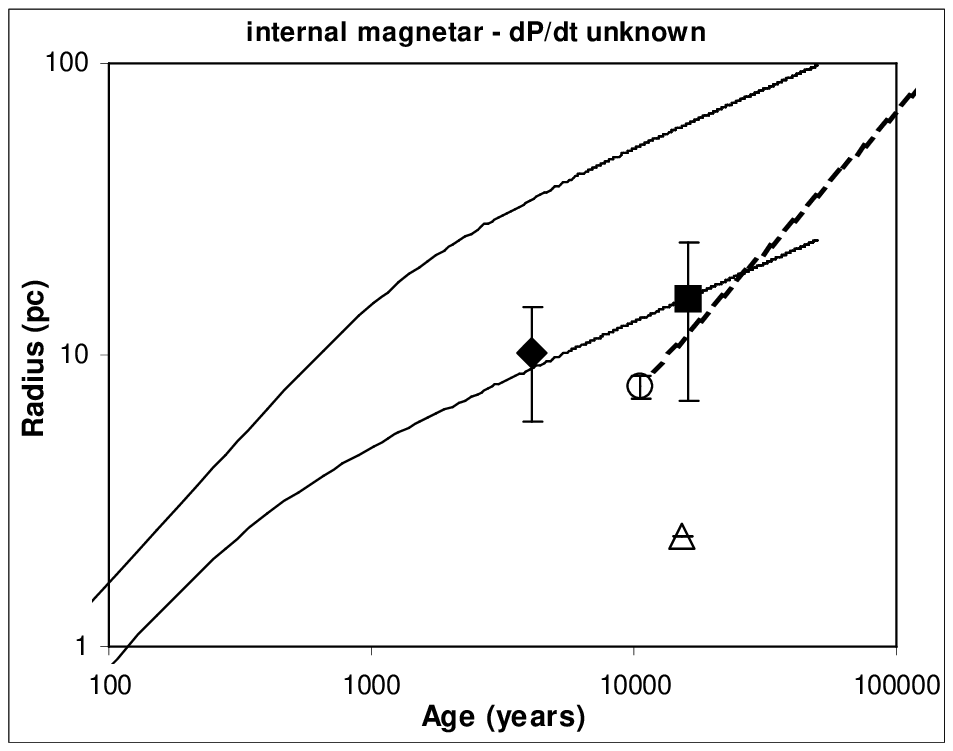}
\caption{Comparison of radius and ages of the proposed associations to the
expansion model with energy injection, for objects without estimated $\dot P$. The solid 
lines represent the evolution according to our model. Markers 
are placed according to the median of estimated SNR ages and radii. Error bars
indicate ranges of SNR ages and radii. The dotted line is the
approximate end of Sedov-Taylor phase (including energy injection). Filled
markers represent likely associations, and open markers represent the unlikely
ones (see text). The markers are: open triangle (SGR 1627-41), open circle (SGR 0526-66),
filled diamond (AXP 1845-0258), filled square (SGR 1627-41). \label{f-unknown}}
\end{figure}

\clearpage 

\end{document}